\documentclass[12pt]{iopart}
\usepackage{graphicx}
\usepackage{amssymb}
\usepackage{hyperref}
\usepackage{url}
\usepackage{float}
\usepackage{enumerate}
\usepackage{color}
\usepackage{amssymb}
\usepackage{amstext}

\newcommand{\ket}[1]{\left\vert #1\,\right\rangle}
\newcommand{\bra}[1]{\left\langle #1\,\right\vert}

\begin{document}

\setlength\abovedisplayskip{5pt}
\setlength\belowdisplayskip{5pt}
\title[Process of equilibration in many-body isolated systems]{Process of equilibration in many-body isolated systems: Diagonal versus thermodynamic entropy}
\author{Samy Mailoud}
\address{Instituto de F\'{i}sica, Benem\'{e}rita Universidad Aut\'{o}noma
  de Puebla, Apartado Postal J-48, Puebla 72570, Mexico}
\ead{samy.mailoud90@gmail.com}
\author{Fausto Borgonovi}
\address{Dipartimento di Matematica e
  Fisica and Interdisciplinary Laboratories for Advanced Materials Physics,
  Universit\`a Cattolica, via Musei 41, 25121 Brescia, Italy}
\address{Istituto Nazionale di Fisica Nucleare,  Sezione di Pavia,
  via Bassi 6, I-27100,  Pavia, Italy}
\author{Felix M. Izrailev}
\address{Instituto de F\'{i}sica, Benem\'{e}rita Universidad Aut\'{o}noma
  de Puebla, Apartado Postal J-48, Puebla 72570, Mexico}
\address{Department of Physics and Astronomy, Michigan State University, E. Lansing, Michigan 48824-1321, USA}

\date{\today}

\begin{abstract}

As recently manifested \cite{BIS19a}, the quench dynamics of isolated quantum systems consisting of a finite number of particles, is characterized by an exponential spreading of wave packets in the many-body Hilbert space. This happens when the inter-particle interaction is strong enough, thus resulting in a chaotic structure of the many-body eigenstates considered in an unperturbed basis. The semi-analytical approach used here, allows one to estimate the rate of the exponential growth as well as the relaxation time, after which the equilibration (thermalization) emerges. The key ingredient parameter in the description of this process is the width $\Gamma$ of the Local Density of States (LDoS) defined by the  initially excited state, the number of particles and the interaction strength. In this paper we show that apart from the meaning of $\Gamma$ as the decay rate of survival probability, the width of the LDoS is directly related to the diagonal entropy and the latter can be linked to the thermodynamic entropy of a system  equilibrium state  emerging  after the complete relaxation. The  analytical expression relating the two entropies is derived phenomenologically and numerically confirmed  in a model of bosons with random two-body interaction, as well as in a deterministic model which becomes  completely integrable in the continuous limit. 
\end{abstract}\begin{flushleft}

\end{flushleft}


\section{Introduction}

The  problem of thermalization in isolated quantum systems remains a hot topic in the field of modern statistical mechanics. It has  been shown since long that thermalization can emerge without the presence of a heat bath, even if the number of interacting particles is small \cite{HZB95,ZBFH96,FI97,Z96,BGIC98,GGF99}. One of the main problems in this field is to establish the conditions under which a given system manifests strong statistical properties, such as the relaxation to equilibrium. Due to a remarkable progress in the study of this and related problems, much is already understood  in  theoretical and  numerical approaches (see \cite{RDO08,PSSV11,DKPR16,LL16,AABS19} and references therein) as well as in experimental studies of interacting particles in optical traps \cite{GMHB02,To12,Go12,Ko16,NH15}, even if many basic problems still need further intensive efforts. 

As was shown in Ref.~\cite{our}, the mechanism for the onset of  statistical behavior for a quantum isolated system is the chaotic structure of the many-body eigenstates in a given basis defined in the absence of the inter-particle interaction. As for the properties of eigenvalues (such as the Wigner-Dyson type of the nearest-neighbor level spacings distribution), it was shown~\cite{SBI12,SBI12a} that they have a little impact on the global statistical properties of the wave packet dynamics. 

The concept of chaotic eigenstates originates from Random Matrix Theory (RMT), which was suggested by Wigner in order to explain local properties of energy spectra of heavy nuclei, observed experimentally \cite{Wigner,Dyson}. Being unable to describe global properties of energy spectra of complex physical systems, the random matrices turned out to be effective models for the description of the local statistical properties of spectra, that were predicted and later confirmed experimentally to be universal. Since long time, the RMT has served as the theory of quantum chaos with strong chaotic properties. 

Next steps in the mathematical study of many-body chaotic systems were performed by taking into account  the typical 
two-body nature of the inter-particle interaction. As a result, a new kind of random matrix model, closer to the physical realm than the RMT and  known as the Two-Body Random Interaction (TBRI) model 
has been invented \cite{FW70,BF71,brody}. In contrast with the standard RMT, it depends on additional physical parameters, such as the number of interacting particles and the strength of inter-particle interaction. In these matrices a complete randomness is embedded into the two-body matrix elements, from which the many-body matrix elements are constructed. 

A distinctive property of the TBRI matrices (in application to both Fermi and Bose particles) is that they are non-ergodic in the sense that the averaging inside a matrix is not equal to the ensemble averaging, even for a very large  matrix size\cite{BW03}.
Moreover, these matrices are banded-like, sparsed and  non-invariant under rotations so that  all their properties are related to a specifically given unperturbed basis. For this reason, rigorous analytical analysis (especially, for a finite number of particles) is strongly restricted and many results can be  obtained only numerically. 

To date, the TBRI matrices are extensively studied, however, mainly in what concerns the properties of energy spectra and structure of eigenstates (see for instance  \cite{BW03,CKP12,kota-book,KC18,VK19} and references therein).  As for the related time evolution, a close attention to this problem was recently surged ahed  by the  increasing interest to the problem of scrambling, understood as the loss of information in the process of equilibration and thermalization.  Further progress in random matrix theories was achieved due to remarkable results based on the Sachdev-Ye-Kitaev (SYK) model \cite{SYK-1,SYK-2}, which can be considered as a variant of the TBRI model (see, for instance, Ref. \cite{JV19}). The SYK model has attracted much attention in recent years, widely accepted as an effective model for two-dimensional gravity, also in application to black holes \cite{SS14,CHLY17,BDGL18}. For the latter problem, a particular interest has been given in the spreading and disappearing of information  in  the process of thermalization. 

As is known, chaos in classical systems is originated from the exponential sensitivity of motion with respect to small perturbations. As a result, the distance in the phase space between two close trajectories increases, in average, exponentially fast and the rate of such an instability is given by the largest Lyapunov exponent $\lambda$. In quantum systems, this mechanism is absent due to the linear nature of the Schrodinger equation. Although for quantum systems with a strongly chaotic classical limit, one can observe a complete correspondence to the classical behavior, the time scale $t \sim 1/\lambda$ on which it happens, was found to be dramatically short due to the fast spreading of the wave function.
As a result, the chaotic behavior of quantum observables is suppressed in time. 

As for quantum systems without classical limit, even in the presence of strong disordered potentials,  observation of exponential instability in the  quantum motion was questioned for a long time. However, recent remarkable progresses have led to the discovery of the out-of-time-order correlators (OTOC), which are four-point correlation functions with a specific time ordering. Extensive studies have manifested the effectiveness of the OTOC in application to various physical systems \cite{LO69,MSS16,RGG17,S18}. 

It is widely believed that  OTOCs can solve the problem of thermalization, however, this is not obvious since their exponential growth in time  is bounded by a time scale which cannot be associated with the complete relaxation  to equilibrium \cite{Rossini2010}. Indeed, typical applications of  OTOC are mainly restricted by {\it local} observables, in contrast with the very point that thermalization is a {\it global} concept. The situation reminds that which was thoroughly discussed in the  early stage of the setting up of the classical chaos theory. Specifically, for some time it was believed that correlation functions should typically manifest an exponential time decrease. Unexpectedly, it was found later that realistic physical systems, even if strongly chaotic,  are typically characterized by a  quite slow decrease of correlations.
Also, an infinite number of correlations functions can be defined and their time behavior can be in principle very different. Thus, even if serving as a good test for the instability of quantum many-body systems, the direct relevance of the OTOC to the time scale of  thermalization remains questionable. 

One of the first attempts to relate the OTOC to the long-time dynamics in many-body systems has been  performed in Ref~\cite{BIS19b}. Specifically, it was asked whether  OTOC can describe the  exponential long-time growth of the effective number $N_{pc}$ of components of the wave function, in the process of a complete equilibration. This number can be estimated via the {\it participation ratio} provided the many-body eigenstates can be considered as strongly chaotic. An important result of this study is that there are two characteristic time scales, one of which is directly related to the survival  probability and can be defined in terms of the Lyapunov exponent, therefore, in terms of the OTOC. However, a complete thermalization is described by the excitation flow along a kind of network created by the many-body states. An application of  OTOC to such type of dynamics \cite{BIS19b} has shown that $N_{pc}$ can be presented as a set of the OTOCs, each of them describing an excitation on a specific time scale, due to standard perturbation theory. For a finite number of many-body states, this process terminates when all   states are excited, which create an energy shell in the Hilbert space (for details see \cite{our}). 

Numerical data obtained for the  TBRI model  with a finite number of bosons occupying a number of single-particle energy levels\cite{BIS19a}, as well as for  models of a finite number of interacting spins-1/2 in a finite length chain\cite{SBI12, SBI12a}, clearly demonstrated that, in presence of chaotic eigenstates,   the global time behavior of the systems is very similar. One has to note that even if one of the spin models is integrable, with a Poisson-like level space distribution, this does not influence the quench dynamics. These results confirm the prediction that for many-body systems the type of energy level fluctuations is less important in comparison with the chaotic  structure of the many-body states. 

In this paper we continue the study of the quench dynamics, paying attention to the  new question of the relevance between the diagonal entropy related to an initially excited state, and the thermodynamic entropy emerging in the process of thermalization. We show numerically and describe semi-analytically that there is a one-to-one correspondence between them, with some corrections due to different size between  unperturbed and perturbed energy spectra. This remarkable result holds both for the TBRI model with finite number of bosons, and for the model originated from the celebrated Lieb-Liniger (LL) model \cite{LL63,L63,Girardeu}. The latter model, which has no random parameters, was proved to be integrable with the use of the Bethe ansatz. 

The LL model describes one-dimensional (1D) bosons  on a circle interacting  with a two-body point-like interaction.  It belongs to a peculiar class of quantum integrable models solved by the Bethe ansatz \cite{Bethe, Korepin}; in particular, it possible to show that it has an infinite number of conserved quantities. Apart from the theoretical interest, this model is important in view of various experiments with atomic gases\cite{Go01,Po04,KWW04}. For a weak inter-particle interaction the LL model can be described in the mean-field (MF) approximation. Contrarily, for a  strong interaction, the 1D atomic gas enters the so-called Tonks-Girardeau (TG) regime in which the density of the interacting bosons becomes identical to that of non-interacting fermions (keeping, however, the bosonic symmetry for the wave function) \cite{Girardeu}. The crossover from one regime to the other is governed by the ratio $n/g$ between the boson density $n$ and the interaction strength  $g$. The latter constant is inversely proportional to the 1D inter-atomic scattering length and can be experimentally tuned with the use of the Feshbach resonance (see, for example, \cite{Olshanii} and references therein). Specifically, the MF regime occurs for $n/g \gg 1$ and the TG regime emerges for $n/g \ll 1$ \cite{Girardeu}.

In our numerical study we consider a finite many-body Hilbert space by fixing the total number of momentum states and the number $N$ of interacting bosons. This truncated Lieb-Liniger model (TLL) allows one to correctly obtain both the eigenvalues and many-body eigenstates of the original LL model, in a given range of energy spectrum \cite{new}. This method (truncation of the infinite spectrum) can be considered as the complimentary one to the recently suggested way \cite{RKC19}, according to which a finite number of eigenstates involving into the quench dynamics is used. 

\section{Quench dynamics}
Below, we study the quench dynamics in two models described by the Hamiltonians written as
\begin{equation}
  H= H_0+V.
\label{ham_gen}
\end{equation}
Here the first term $H_0$ describes $N$ non-interacting identical bosons occupying $M$ single-particle levels specified by the energies $\epsilon_s$,
\begin{equation}
  H_0 =  \sum \epsilon_s \, \hat{n} _s ,
\label{H0}
\end{equation}
while the second term $V$ stands for the two-body interaction between the particles,
\begin{equation}
  V = \sum V_{s_1 s_2 s_3 s_4} \, a^\dag_{s_1} a^\dag_{s_2} a_{s_3} a_{s_4}.
\label{HV}
\end{equation}
Here $\hat{n}_s = \hat{a}^\dagger_s\hat{a}_s$ is the number of particles in the corresponding $s-$level, with $\hat{a}^\dagger_s$ and $\hat{a}_s$ as the creation/annihilation operators acting on the $s$-th  level, and the two-body matrix elements $ V_{s_1 s_2 s_3 s_4} $ specify the type and strength of the inter-particle interaction. The interaction conserves the number of bosons and connects many-body states that differ by the exchange of at most two particles. Due to the two-body nature of the interaction, the matrix elements $H_{ij} = \langle i|H|j\rangle $ are non-zero only when the two unperturbed many-body basis states $\ket{i}$ and $\ket{j} $ have single-particle occupations which differ by no more than two units. This means that the Hamiltonian matrix is sparse which is a common  property of realistic many-body systems. 

Our aim is to compare the dynamical properties of two models:
\begin{enumerate}[i.]
	\item the TBRI model with completely random and independent values of $ V_{s_1 s_2 s_3 s_4} $
	\item the TLL model which is deterministic (without random parameters) and integrable in the limit $M \rightarrow \infty$.
\end{enumerate}
 
The unperturbed many-body eigenstates $\ket{k} $ of 
$$H_0= \sum_k {\cal E}_k \ket{k} \bra{k} $$
  are obtained by creating $N$ bosons in $M$ single-particles energy levels, so that $|k\rangle = a^\dagger_{s_1}...a^\dagger_{s_N} |0\rangle$, where $1 \leq s_1,..,s_N \leq M$. In this way, we determine the unperturbed many-particles basis, in which the Hamiltonian is a diagonal matrix. As a result, the many-body eigenstates $\ket{\alpha}$ of the total Hamiltonian $H$ are represented in terms of  the basis states $|k\rangle$ as 
\begin{equation}
\ket{\alpha} = \sum_k C_k^{\alpha} \ket{k} ,
\label{alpha}
\end{equation}
where $C_k^{\alpha}$ are obtained by exact numerical diagonalization. 

After specifying the unperturbed basis $\ket{k}$, one can study the wave packet dynamics in this basis, after switching on the interaction $V$. All analytical and numerical results refer to the situation when initially the system is prepared in a particular unperturbed state, 
\begin{equation}
\ket{k_0}  = \sum_\alpha C_{k_0}^\alpha \ket{\alpha}.
\end{equation}
Thus, our main interest is to explore the evolution of wave packets in the many-particle basis of $H_0$, starting from an unperturbed state $\ket{j_0}$. 
To do this, we focus on the time dependence of the effective number $N_{pc}(t)$ of principal components of the wave function,
\begin{equation}
N_{pc}(t) \equiv  \left( \sum_k |\langle k |e^{-iHt} |j_0\rangle |^4 \right)^{-1},
\label{Npc} 
\end{equation}
known in  literature as the {\it participation ratio}.

In terms of eigenvalues and many-body eigenstates of the Hamiltonian $H$ the participation ratio can be presented as
\begin{equation}
\label{s-ipr}
 N_{pc}(t) = \left\{ \sum \limits_{k} \left[P_{k}^d + P_{k}^f (t)\right]^2  \right\}^{-1},
\end{equation}
where 
\begin{equation}
P_{k}^d=\sum_{\alpha}  |C_{k_0}^\alpha|^2 |C_{k}^\alpha|^2
\label{Pkd}
\end{equation}
and
\begin{equation} 
P_{k}^f (t)=\sum_{\alpha\ne \beta}  C_{k_0}^\alpha C_{k}^{\alpha \ast} C_{k_0}^\beta C_{k}^{\beta \ast} e^{-i(E^\beta-E^\alpha )t}
\label{Pkf}
\end{equation} 
are the diagonal and off-diagonal parts of   
\begin{equation}
P_k(t) = |\langle k |\psi (t) \rangle |^2 = \sum_{\alpha,\beta}  C_{k_0}^{\alpha \ast} C_{k}^{\alpha } C_{k_0}^\beta C_{k}^{\beta \ast} e^{-i(E^\beta-E^\alpha )t} .
\label{Pkt}
\end{equation}

As was recently shown for different models with chaotic behavior~\cite{BIS19a}, the number $N_{pc}$ increases exponentially fast in time, provided the eigenstates involved in the dynamics, are strongly chaotic. After some relaxation time $t_s$, the value of $N_{pc}$ fluctuates around the saturation value due to the complete filling of a portion of the total Hilbert space, called energy shell. Such a wave packet dynamics occurs when the many-body eigenstates of $H$ are fully delocalized in the energy shell. This scenario explains the basic properties of the quench dynamics, before and after the saturation. 

\section{Semi-analytical approach}
The semi-analytical approach developed in \cite{FI97} for TBRI matrices with an infinite number of interacting Fermi-particles and modified in \cite{BIS19a} for finite Bose-systems, allows one to obtain simple estimates for two important characteristics of the quench dynamics. The first characteristic is the rate of the exponential growth for the $N_{pc}(t)$. The second characteristic is the time scale $t_s$ on which the exponential growth  of $N_{pc}(t)$ occurs.



The two above characteristics can be estimated with the use of the semi-analytical approach originally developed in \cite{FI01b} for an infinite number of particles. To start with, we write down an infinite set of the probability conservation equations, used for the description of the probability flow $W(t)$ in the many-body Hilbert space of $H_0$,
\begin{equation}
\begin{array}{lll}
\displaystyle \frac{dW_0}{dt} &= -\Gamma W_0 , \\
&\\
\displaystyle \frac{dW_1}{dt} &= -\Gamma W_1 +\Gamma W_0 ,\\
&\\
\displaystyle \frac{dW_2}{dt} &= -\Gamma W_2 +\Gamma W_1  ,\\
&\\
...
\label{casmod}
\end{array}
\end{equation}
Here $W_0(t) \equiv {P_{k_0}(t)}$ is the standard survival probability which is the probability to find the system in the initial state $\ket{k_0}$ at the time $t$. It is characterized by the decay width $\Gamma$ according to which the probability flows to those unperturbed states which are directly coupled to the initial one via the two-body interaction $V$. These states define the set ${\cal M}_1$ from which the probability flows onto another set ${\cal M}_2$ in the second order of perturbation theory. Similarly, in the next order of perturbation theory the probability flows to the set ${\cal M}_3$ and so on, thus describing the spread of the probability on a "tree" created by an infinite number of subsets $W_k(t)$. In Ref.\cite{FI01b} it was shown that the above equations can be solved exactly:
\begin{equation}
W_k(t)= \frac{(\Gamma t)^k}{k!} W_0(t) 
\label{solution}
\end{equation}
with $W_0(t) = \exp (-\Gamma t)$. 

The above solution (\ref{solution}) allows one to derive the expressions for various observables. In particular, for large value of the {\it connectivity} ${\cal K}$, in \cite{FI01b} the following approximate expression has been obtained,
\begin{equation}
N_{pc}(t)= \exp\big[2\Gamma(1-\frac{1}{\sqrt{\cal K}})t]. 
\label{NPC}
\end{equation}
Here $\cal K$ is the number of non-zero matrix elements of $H$ along a particular line associated with the initial state $\ket{k_0}$.
It should be stressed that the analogy with an infinitely large "tree" is correct in the thermodynamic limit only.  For a finite Hilbert space, as in our case, the  number of the effective subsets $W_k(t)$ is finite. As a result, as $t \rightarrow \infty$ each of the terms $W_k(t)$ converges to a non-zero value.  Moreover, the set of equations (\ref{casmod}) is practically restricted by a few sets (in our simulations, by $W_1(t)$ and $W_2(t)$, since the number of elements in ${\cal M}_2$  is already of the same order as the dimension of the total Hilbert space).
For the case of a small number of subsets, in Ref.~\cite{BIS19a} the equations (\ref{casmod}) have been modified and used to explicitly obtain an extremely detailed  quench dynamics. In this case one can show that, on an intermediate time scale,  
\begin{equation}
N_{pc}(t) \approx \exp (2\Gamma t), 
\label{NPC-short}
\end{equation}
which coincides with Eq.~(\ref{NPC}) for large ${\cal K}$ values. 


From the above analysis one can see that the key parameter in the quench dynamics is the width of the LDoS,  $\Gamma$.  The Local Density of States is defined by,
\begin{equation}
F_{k_0} (E) =\sum_{\alpha}  | C_{k_0}^{\alpha}|^2 \delta (E - E^{\alpha }) , 
\label{ldos} 
\end{equation}
where $\ket{k_0}$ is an eigenstate of $H_0$. Thus, it is obtained by the projection of the initial state $\ket{k_0}$ onto the exact eigenstates of $H$. The concept of LDoS is extremely important in the analysis of the dynamical properties of many-body systems. For example, the Fourier transform of the LDoS determines the return probability of an initially excited many-body state, and it is effectively used in the study of fidelity in many applications. The inverse width $1/\Gamma$ of LDoS gives the  characteristic time scale, which is associated with the depletion of the initial state, thus representing an early stage of thermalization only \cite{BIS19b}. Initially introduced in atomic \cite{R33} and widely used in nuclear physics \cite{BM69}, it also serves as an important characteristic in other physical applications. As shown in many different  papers (see for instance \cite{I01} and references therein), for systems with well defined classical limits, a classical analog of the LDoS can be defined and directly computed from the Hamilton equations of motion. 

For isolated systems of interacting particles, described by a  Hamiltonian $H=H_0 + V$, the form of LDoS strongly changes on increasing  the perturbation strength $V$~\cite{our}. If for a weak, but not negligible, interaction the LDoS is typically a Lorenzian (apart from the tails which are due to the finite width of the energy spectrum), for a strong interaction (when $V \approx H_0$) its form becomes close to a  Gaussian. Correspondingly, the width   $\Gamma$ of the LDoS can be estimated either using the Fermi golden rule, $\Gamma \approx 2\pi V^2 \rho_f$ where $\rho_f$ is the density of the many-body states directly connected by $V$, or by the square root of the variance of LDoS, $\sigma = \sqrt{\sum H_{ij}^2}$ for $i\neq j$. When studying the TBRI model, this crossover was found to serve as the condition for the onset of strong quantum chaos,  defined in terms of a pseudo-random structure of many-body eigenstates. For this reason, instead of $\Gamma$ in our case one can use $\sigma$ since the latter is much easier to estimate than the former. Thus, when comparing our data in Fig.\ref{fig:pr} with the predicted exponential dependence (\ref{NPC-short}), we use the following expression:
\begin{equation}
\Gamma^2 =   \sum_{k\ne j_0} H^2_{k,j_0}  .
\label{eq:gamma} 
\end{equation}
As will be shown below, the simple expression (\ref{NPC-short}) nicely corresponds to numerical data demonstrating the exponential increase of $N_{pc}$ in time. 

Now let us discuss another important characteristic of the relaxation for finite systems, namely, the time scale over which the exponential growth of $N_{pc}$ lasts. To do that, we have to estimate the saturation value $\overline{N_{pc}^{\infty}}$ of $N_{pc}$ after the relaxation of the system to equilibrium. This value, can be obtained by the time average performed after the  saturation time $t_s$,
\begin{equation}
\label{eq:npcinfty}
{ [ \ \overline {N_{pc}^{\infty}} \ ]^{-1} } = \lim_{T\to \infty} \frac{1}{T} \int_0^T \ dt \ [N_{pc}(t)]^{-1}.
\end{equation}
An analytical expression, assuming non-degenerate energy levels, in terms of the eigenstates can be written as,
\begin{equation}
\label{t-ipr}
\overline{N_{pc}^{\infty}} =  \left[ 2 \sum_k (P_{k}^d)^2  -  \sum_{\alpha}  | C_{k_0}^\alpha |^4 \sum_k |C_{k}^\alpha|^4 \right]^{-1}.
\end{equation}
This expression determines the total number of unperturbed many-body states inside the energy shell, excited in the process of equilibration. It was shown \cite{BIS19a} that for strongly chaotic eigenstates the first term in Eq.(\ref{t-ipr}) is much larger than the second one. With further simplifications one can obtain the following estimate:
\begin{equation}
\overline{N_{pc}^{\infty}} \approx f(\eta) \mathcal {D} ,
\label{estimate-1} 
\end{equation}
where $\mathcal {D}$ is the size of the Hamiltonian matrix  and $f$ is a function of $\eta = M/N$ (here it is assumed that both $M$ and $N$ are much larger than 1). As one can see, the meaning of $\overline{N_{pc}^{\infty}}$ is simply the partial size of the many-body  Hilbert space in which the dynamics occurs. 
The critical time $t_s$ determining the onset of the saturation, can be estimated from the relation,
\begin{equation}
\label{t-s1}
\exp (2\Gamma t_s) \approx \overline{N_{pc}^{\infty}}.
\end{equation}
Assuming a Gaussian shape for both the density of states and the LDoS, $M \simeq 2 N$, and taking the maximal value of the function $f(\eta)$, one gets,
\begin{equation}
\label{t-s2}
t_s   \approx \frac{N}{ \Gamma} .
\end{equation}

This is one of the important results, obtained in the frame of the discussed approach. As one can see, the characteristic time $t_s$ is $N$ times larger than the time $t_{\Gamma} \approx 1/\Gamma$ describing an early decrease of the return probability. The key point is that the time $t_{\Gamma}$ has to be associated only with an initial process towards the true thermalization. In contrast, the latter emerges when the flow of probability fills {\it all} the subsets $W_k$  that create the energy shell available in the thermalization process. This result can have important implications for addressing other issues such as the scrambling of information and the quantum butterfly effect.

\section{Thermodynamic entropy versus diagonal entropy}
Now let us discuss an important relation which has been discovered by analyzing the quench dynamics leading to the thermalization. This relation links two entropies, $S_{th}$ and $S_{diag}$. Here
\begin{equation}
S_{th} = \ln {\cal V}(E)
\label{s-th}
\end{equation}
 is the thermodynamic entropy characterizing the system {\it after} its relaxation to equilibrium and ${\cal V}$ is the ``volume'' associated with the part of the Hilbert space, occupied by the wave function. It can be estimated as
\begin{equation}
{\cal V}(E) \simeq \overline{N_{pc}^{\infty}} \delta_0,
\label{vol1}
\end{equation} 
where $\delta_0=\Delta_{H_0}/{\cal D}$ is the unperturbed energy spacing, $\Delta_{H_0}$ is the effective width of the energy spectrum of $H_0$ and ${\cal D}$ is the dimension of the many-body Hilbert space. 

As for the diagonal entropy $S_{diag}$,  discussed in view of its relation to the Von Neumann entropy \cite{P11}, it is given by,
\begin{equation}
\label{s-diag}
S_{diag}  = -\sum_{\alpha }  | C_{k_0}^\alpha |^2  \ln | C_{k_0}^\alpha |^2.   
\end{equation}
Note that the diagonal entropy is the Shannon entropy of the set of probabilities $w_{k_0} (E^\alpha)  =| C_{k_0}^{\alpha}|^2$ obtained by the projection of the  unperturbed state $\ket{k_0}$  of $H_0$ onto the exact states of $H$. With the Shannon entropy we can built the entropic localization length 
\begin{equation}
\label{ent-ell}
\ell_H = \exp \left( S_{diag} \right),
\end{equation}
 giving the number of exact eigenstates of  $H$ excited by the initial unperturbed state \cite{FI01b}. Thus, the volume occupied by the initial state is  ${\cal V}(E) \simeq \ell_H \delta$, where $\delta$ is the energy spacing estimated as $\delta \simeq  \Delta_H/{\cal D}$. Comparing the two volumes, $ \langle N_{pc}\rangle \Delta_{H_0}$
and  $\ell_H \Delta_{H}$, we arrive at the following relation: 
\begin{equation}
S_{th} = S_{diag} + \ln (\Delta_H / \Delta_{H_0}) ,
\label{relation}
\end{equation}
where $\Delta_H$ and $\Delta_{H_{0}}$ are the widths of the energy spectra of $H$ and $H_0$, respectively.  A similar correction due to the difference between $\Delta_H$ and  $\Delta_{H_0}$ also appeared in other context~\cite{BI00}. 

\subsection{Two-body random interaction model}
\begin{figure}[t]
\centering
\includegraphics[width=8cm]{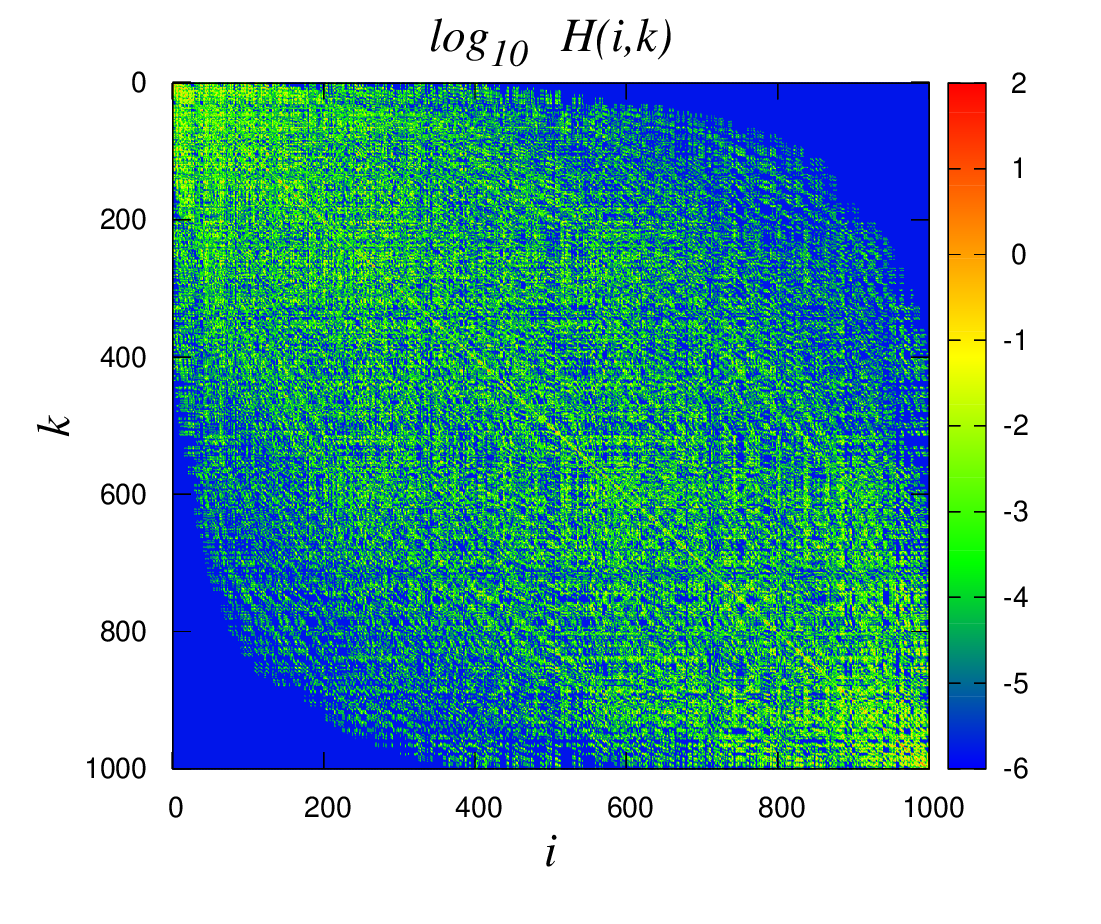}
\caption{Global structure of the Hamiltonian matrix for the TBRI model with $N=4$ particles, $M =11$ momentum states and interaction strength $v=0.4$.  }
\label{fig:TBREmatr}
 \end{figure}

 The TBRI model is characterized by the Hamiltonian of $N$ interacting bosons occupying $M$ single-particle levels, 
\begin{equation}
  H= \sum \epsilon_s \, a^\dag_s a_s  +
 \sum V_{s_1 s_2 s_3 s_4} \, a^\dag_{s_1} a^\dag_{s_2} a_{s_3} a_{s_4} ,
\label{ham}
\end{equation}
in which the single-particle levels are specified by random energies $\epsilon_s$ with  mean spacing $\langle \epsilon_s- \epsilon_{s-1} \rangle = 1 $ setting the energy scale. This model belongs to the class of random models since the two-body matrix elements $ V_{s_1 s_2 s_3 s_4} $ are random Gaussian entries with zero mean and variance $v^2$. Originally suggested to describe systems with Fermi particles, recently the TBRI model has been applied to randomly interacting bosons~\cite{BW03,CKP12,KC18,VK19}. As mentioned above, the Hamiltonian matrix $H$ has a band-like form  and has many zero off-diagonal elements, due to the two-body nature of the inter-particle interaction $V$. It is also worthwhile to mention that being constructed from random entries $V_{s_1 s_2 s_3 s_4} $, the many-body matrix elements are slightly correlated since some of them are constructed from the same two-body matrix elements. How important are these tiny correlations, have been studied analytically and numerically in Ref.\cite{FGI96}. The typical structure of the Hamiltonian matrix $H$ is shown in Fig.\ref{fig:TBREmatr}. 

The wave packet dynamics in the unperturbed basis $\ket{k}$ after switching on the interaction $V$ was numerically analyzed after finding all many-body eigenstates of $H$ and their energy eigenvalues. As one can see, the numerical study is strongly restricted by the number of particles $N$ and the numbers $M$ of single-particle states since the total size of the Hamiltonian matrix increases exponentially with both $N$ and $M$. In our study we have mainly considered the case when $M \approx N$, which is akin to the study of one-dimensional lattices with $M$ sites and $N$ Fermi-particles. As is known, in many aspects some of the observables have the same properties when both $N$ and $M$ increase, keeping  the ratio $N/M$  fixed. 
\begin{figure}[t]
	\centering
	\includegraphics[width=8cm]{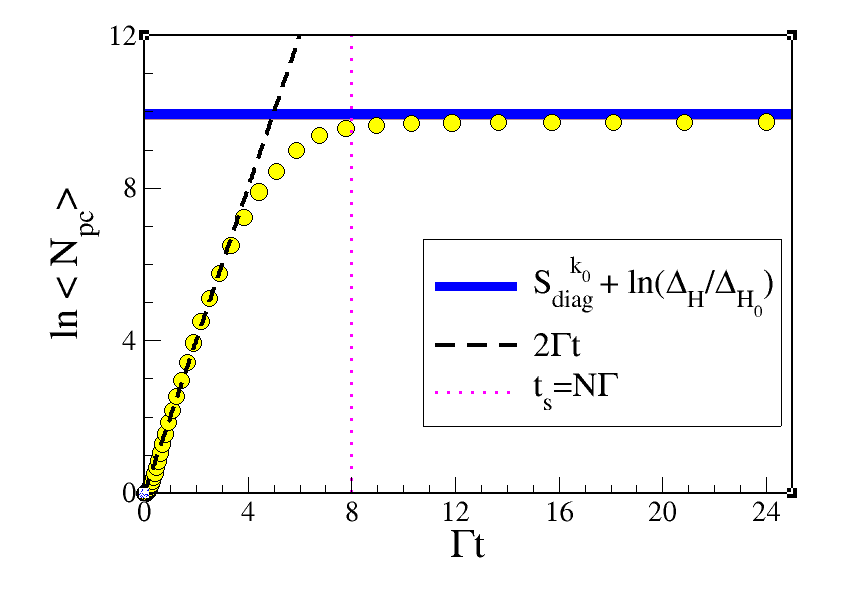}
	\caption{Time dependence of $\ln \langle N_{pc} \rangle$ for the TBRI model. The average has been done over 5 disorder realizations. The initial state $\ket{k_0}$ has all $N=8$ particles in the $5^{th}$ single particle energy level. The total number of available energy levels is $M=11$,  the dimension of the Hilbert space ${\cal D} = 43758$ and the perturbation strength is $v=0.4$. The horizontal blue thick line represents the normalized diagonal entropy $S_{diag}^{k_0}+\ln (\Delta_H/\Delta_{0})$, with the thickness corresponding to one standard deviation due to different realizations of the disorder. The dashed line corresponds to $2\Gamma t$, where $\Gamma$ is the  width of the LDoS associated with the initial unperturbed state and averaged over the 5 disordered realizations.. On horizontal axis we used the dimensionless time $\Gamma t$. Vertical dotted line represents the estimate $\Gamma t_s = N $, where $N=8$ is the number of particles. 
}
\label{fig:tbreqd}
\end{figure}

Our interest is in the time dependence of the number of principal components $N_{pc} (t)$ in the wave function considered in the unperturbed basis of $H_0$. As we discussed in the previous sections, the system is initially prepared in a particular unperturbed state $\ket{k_0}$, which determines the quench dynamics. Having all eigenstates and eigenvalues, we obtain the number $N_{pc} (t)$ as is shown in Fig.~\ref{fig:tbreqd}. For sake of completeness we also plot in the same figure the curve $2\Gamma t $ corresponding to the analytical prediction obtained beyond an initial time scale on which the perturbation theory is valid. As one can see, this prediction is fully confirmed by numerical data.

Now, in order to check the relation between the diagonal and thermodynamic entropies, one needs to have an accurate estimate of the average value of $N_{pc}$ after the relaxation~\cite{BIS19a}. First of all let us notice that the second term in the r.h.s of Eq.~(\ref{t-ipr}) is roughly $1/{\cal D}$ times smaller than the first one, where ${\cal D} $ is the dimension of the many-body Hilbert space. This can be seen by taking uncorrelated components
$ C_{k}^\alpha  \simeq ({1/\sqrt{\cal D}}) e^{i\xi_{\alpha,k}}$, where $ \xi_{\alpha,k}$ are random numbers. Thus, one gets, 
\begin{equation}
\label{t-ipr1}
 2 \sum_k (P_{k}^d)^2 = 2\sum_{\alpha,\beta, k}  | C_{k_0}^\alpha |^2  |C_{k}^\alpha|^2 | C_{k_0}^\beta |^2  |C_{k}^\beta|^2 
\simeq \frac{{\cal D}^3}{{\cal D}^4} \simeq \frac{1}{\cal D} ,
\end{equation}
while
\begin{equation}
\label{t-ipr2}
\sum_{\alpha,k }  | C_{k_0}^\alpha |^4  |C_{k}^\alpha|^4 \simeq \frac{{{\cal D}^2}}{{\cal D}^4} \simeq \frac{1}{{\cal D}^2}
\end{equation}
As a result (taking the first of the above terms only), we arrive at the estimate,
\begin{equation}
\label{t-ipra}
\left[\, \overline{N_{pc}^{\infty}}\, \right]^{-1} \simeq  2 \sum_k (P_{k}^d)^2 .
\end{equation}
 
\begin{figure}[t]
	\centering
	\includegraphics[width=8cm]{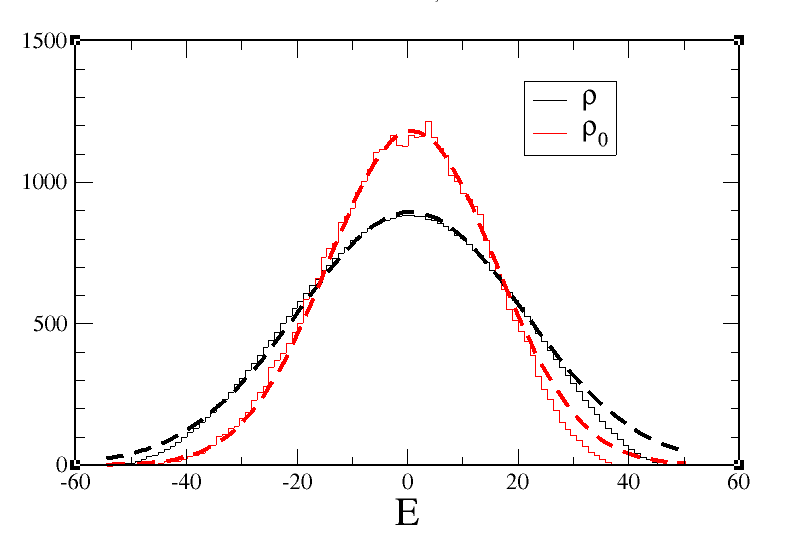}
	\caption{Hystograms stands for perturbed (black) and unpertubed (red) density of states for a system of $N=8$ bosons in $M=11$ single particle energy levels. Interaction strength is $v=0.4$. The effective energy width of perturbed and unperturbed Hamiltonians are : $\Delta_H = 98.78$ and $\Delta_{H_0} = 83.32$. Dashed lines are the Gaussian fit in the central region of the energy spectrum.
}
\label{fig:tbrero}
\end{figure}

We assume a Gaussian shape for (i) the LDoS, (ii) the density of the unperturbed states, and (iii) the density of states of the full Hamiltonian. This is a realistic assumption for chaotic many-body systems with two-body interactions, such as the TBRI model. 

(i) For the LDoS we then have 
\begin{equation}
\label{ldos-c}
F_{k_0}(E) 
 \simeq  \frac{1}{\Gamma\sqrt{2\pi}} 
\exp \left\{ -\frac{(E-{\cal E}_{k_0})^2}{2\Gamma^2} \right\} ,
\end{equation}
where $\Gamma$ is the width of the LDoS and ${\cal E}_{k_0} $ is the energy of the unperturbed state. We also assume  that $\Gamma$ is independent of ${\cal E}_k $. Note that the LDoS is normalized, $\int F_k(E) dE  = 1$. 

(ii) The Gaussian shape  for  the unperturbed density of states $\rho_0(E)$ of width $\sigma_0$ is written as
\begin{equation}
\label{dos0}
\rho_0(E) =     \frac{{\cal D}}{\sigma_0\sqrt{2\pi}} 
\exp \left\{ -\frac{E^2}{2\sigma_0^2} \right\} .
\end{equation} 

(iii) 
The Gaussian density of states, characterized by a width $\sigma$, is such that 
\begin{equation}
\label{dos}
\rho(E) =     \frac{{\cal D}}{\sigma\sqrt{2\pi}} 
\exp \left\{ -\frac{E^2}{2\sigma^2} \right\} ,
\end{equation}
where for simplicity we set the middle of the spectrum at the energy $E=0$.
Both densities of states are normalized to the dimension of the Fock space, 
$$\int  \rho(E) dE = \int  \rho_0(E_0) \ dE_0 ={\cal D}.$$
 Numerical data confirm the Gaussian form of $\rho(E_0)$ and 
$\rho(E)$, see Fig. \ref{fig:tbrero}.  As one can see, due to the perturbation the energy spectrum increases its total width.

The above assumptions imply that in the continuum, one has 
\begin{equation}
\begin{array}{lll}
 &\sum_{\alpha}  | C_{k_0}^\alpha |^2  |C_{k}^\alpha|^2  \simeq  \int \, dE \, \rho(E)^{-1} F_k(E) F_{k_0} (E)= \\
&\\
 &\frac{\sigma^2 \Gamma^{-1}{\cal D}^{-1}}{\sqrt{2\sigma^2-\Gamma^2}}  \exp \left\{ 
-\frac {({\cal E}_{k})^2+({\cal E}_{k_0})^2} {2\Gamma^2}  +   
\frac {({\cal E}_k+{\cal E}_{k_0})^2} {2\Gamma^2(2\sigma^2-\Gamma^2)}\right\} \equiv 
{\cal G}_{k_0} ({\cal E}_k)  
\end{array}
\label{pkd}
\end{equation} 
which is defined only for $2\sigma^2>\Gamma^2$. We can then approximate
\begin{equation}
\label{t-ipra1}
[ \, \overline{N_{pc}^{\infty}} \,] ^{-1} \simeq  2 \sum_k (P_{k}^d)^2  \simeq  2\int \, dE  \, \rho_0(E) {\cal G}_{k_0}^2 (E).
\end{equation}
Taking into account that $\sigma_0^2 = \sigma^2 - \Gamma^2$, Eq.~(\ref{t-ipra1}) gives
\begin{equation}
\label{prf}
 \overline{N_{pc}^{\infty}}  =  C_1  \frac{{\cal D}\Gamma}{\sigma}   e^{-{\cal E}_{k_0}^2/\Gamma^2} 
\end{equation}
with $C_1 = \sqrt{1/2-\Gamma^2/4\sigma^2}$.

\begin{figure}[t]
	\centering
	\includegraphics[width=8cm]{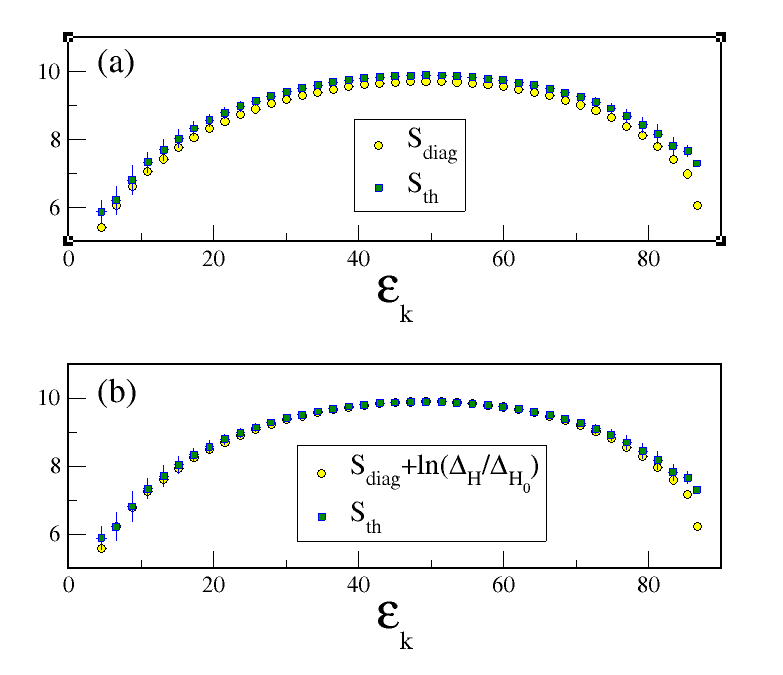}
	\caption{(a) TBRI : Thermodynamic entropy $S_{th} = \ln N_{pc}^{\infty}$ and diagonal entropy $S_{diag}$ {\it vs} the energy of the initial state ${\cal E}_k$. Here we took $N=8$ bosons in $M=11$ single-particle energy levels. Interaction strength $v=0.4$. (b) the same but we added the correction to the diagonal entropy due to the different spectral widths of $H$ and $H_0$.
Average over close energies has been performed. Error bars indicate one standard deviation. 
}
\label{fig:tbrede}
\end{figure}
Let us now estimate the diagonal entropy given in Eq.~\ref{s-diag}, associated with the initial state $\ket{k_0}$ using the same approximations,
\begin{equation}
\label{s-diagc}
S_{diag}   \simeq -\int dE F_{k_0}(E) \ln [F_{k_0}(E)] + \int dE F_{k_0}(E) \ln [\rho(E)].
\end{equation}
Taking only the dominant contribution one gets
\begin{equation}
\label{l-diag}
\ell_{diag} \equiv  \exp\{S_{diag}\}   =  C_2  \frac{{\cal D}\Gamma}{\sigma}   e^{-{\cal E}_{k_0}^2/\Gamma^2} 
\end{equation}
with $C_2 =  \exp[(1-\Gamma^2/\sigma^2)/2]$. Comparing Eq.~(\ref{prf}) with  Eq.~(\ref{l-diag}) one finally gets, 
\begin{equation}
\label{con-ent}
\overline{N_{pc}^{\infty}} = (C_1/C_2) \ell_{diag}.
\end{equation}
This proves that the global functional dependence of both $\overline{N_{pc}^{\infty}}$ and $S_{diag}$ on the initial energy ${\cal E}_{k_0}$ is the same and they turn out to be proportional one to each other. Thus, the data in Fig.~\ref{fig:tbreqd} fully confirm the predicted relation (\ref{relation}). 

In order to check both the consistency of our physical argument and of the relation (\ref{con-ent}) we considered for the whole set of initial energies ${\cal E}_{k_0}$ the two entropies, see Fig.~\ref{fig:tbrede} panel (a).  As one can see they have the same dependence even if they are slightly shifted due to the variation in the size of the energy spectra. Taking into account the correction due to the different spectral widths of $H$ and $H_0$, we can see that the two entropies are pretty much the same, apart from small deviations at the edge of the spectrum where eigenstates are expected to be  non-chaotic, see panel (b) of the same figure. 

\subsection{Truncated LL model} 
 \begin{figure}[t]
\centering
\includegraphics[width=8cm]{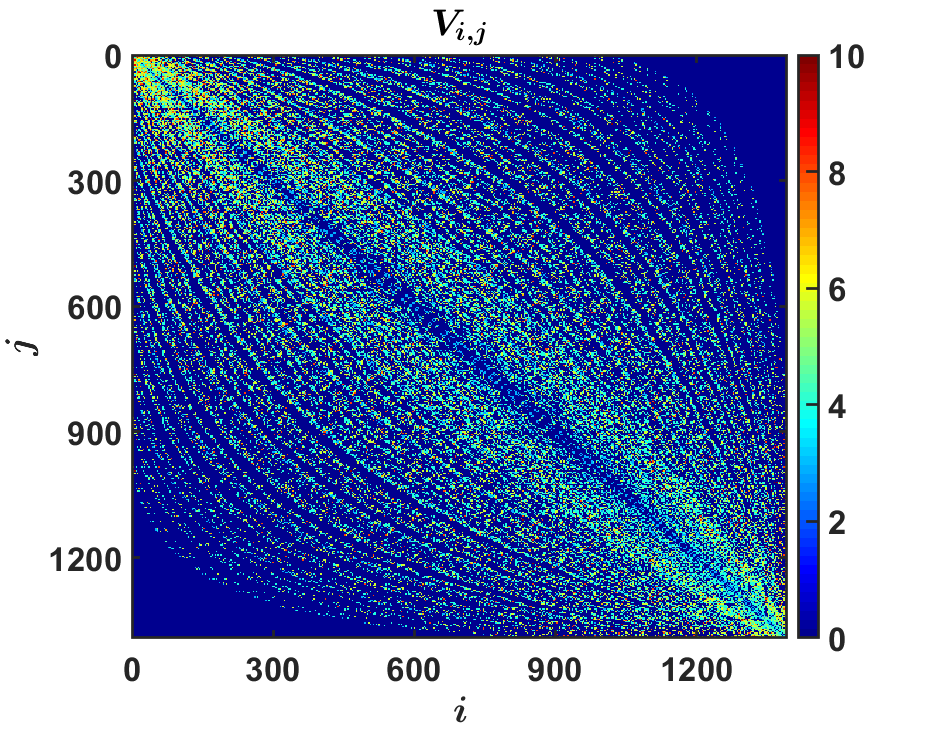}
\caption{TLL model : Structure of the Hamiltonian matrix for a system with $N=5$ particles and $\ell =17$ momentum  states and fixed total momentum   $\mathcal {M}=1$. Here we set $g=1$ and only off-diagonal matrix elements are shown. }
\label{fig:mat}
 \end{figure} 
\begin{figure}[t]
	\centering
	\includegraphics[width=8cm]{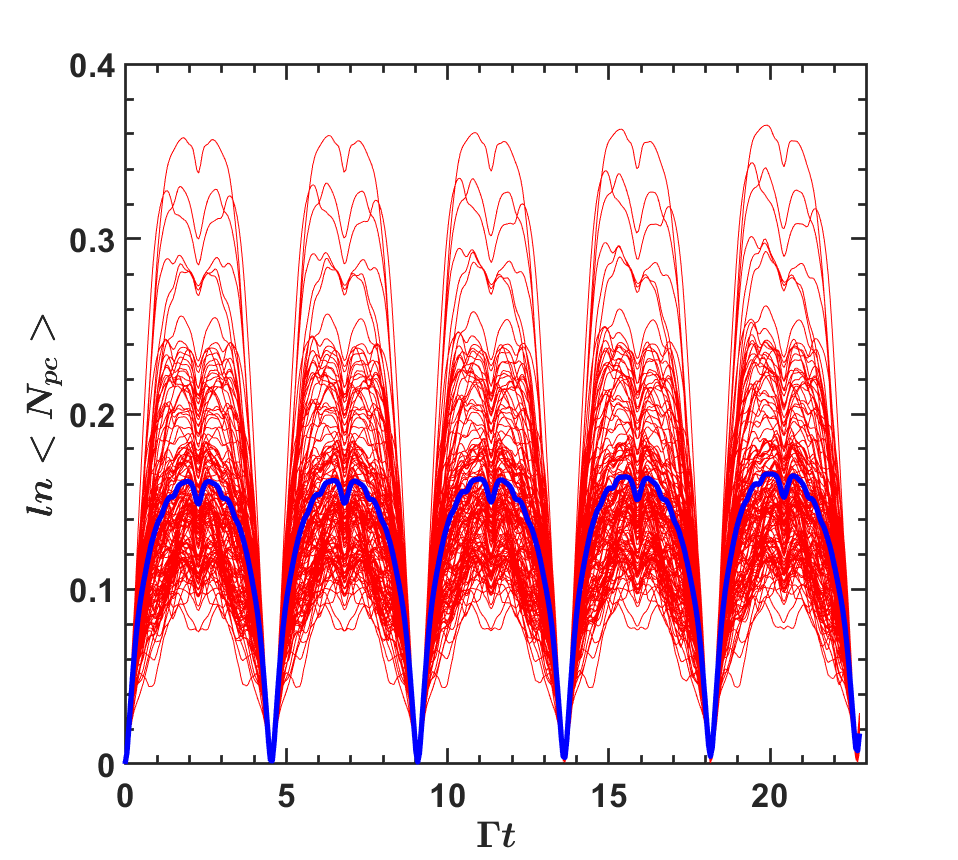}
	\caption{Time dependence of $\ln \langle N_{pc} \rangle$ for the TLL model for weak interaction (MF regime). On x-axis we put the dimensionless time $\Gamma t$ where $\Gamma$ is the  width of the LDoS averaged over the degenerate initial states. The average (blue curve) has been done over all initial degenerate states having the same unperturbed energy $E_0$ close to the band center (with $j_0 \approx 4050$).  Here we considered $N=9$ particles in $\ell=13$ momentum levels, $n/g=10 $, for a fixed total momentum $\mathcal {M}=1$ (the matrix size is $8122$). 
The normalized diagonal entropy $S_{diag}+\ln (\Delta_H/\Delta_{H_0}) = 2.75\pm 1 $ does not correspond to the equilibrium value since for this interaction strength the eigenstates involved in the dynamics are not chaotic. 
}
\label{fig:wpr}
\end{figure}

The Hamiltonian of the Lieb-Liniger model with $N$ bosons occupying a ring of length $L$, in dimensionless units can be written as,  
\begin{equation}
H=\sum_{s} \epsilon_s \hat{n} _s +\frac{g}{L}\sum_{s,q,p,r} \hat{a}^\dagger_s \hat{a}^\dagger_q \hat{a}_p \hat{a}_r \delta (s+q-p-r) .
\label{eq:ham}
\end{equation}
Here $g$ is the parameter determining the strength of interaction between particles, and the single-particle energy levels (for non-interacting bosons) are given by 
$$\epsilon_s = 4 \pi^2 s^2/L^2.$$ The $\delta -$function in Eq.~(\ref{eq:ham}) indicates the momentum conservation of the two-body interaction. Below, the single-particle states $\ket{\phi_s}$ are labeled according to their momentum $s=0, \pm 1, \pm 2,...$. From them, the many-body unperturbed states $\ket{j} =  |... n_{-s}...n_0,..., n_s...\rangle$ have been built, where $n_{s}$ indicates the number of particles in the $s$-th momentum level. In our numerical study, in order to fit physical experiments,  we consider a finite number $N$ of particles occupying a finite number $\ell=2M+1$  of single-particle momentum states. Note that the total number of single-particle energies $\epsilon_s$ is $M+1$ since the states with momentum $\pm s$ are degenerate.  We choose $N$ and $M$ to be approximately the same, in analogy
with the  case considered above for the TBRI model. A rough estimate, $n=N/L \sim g$, for the crossover from the MF to the TG regime in connection with quantum chaos, is discussed in~\cite{BBIS04}.    


The structure of the Hamiltonian matrix at some fixed total momentum value $\cal{M}$ is shown in  Fig.~\ref{fig:mat} and as one can see it is very similar to that presented in Fig.\ref{fig:TBREmatr} for the TRBI model.

\begin{figure}[t]
	\centering
	\includegraphics[width=8cm]{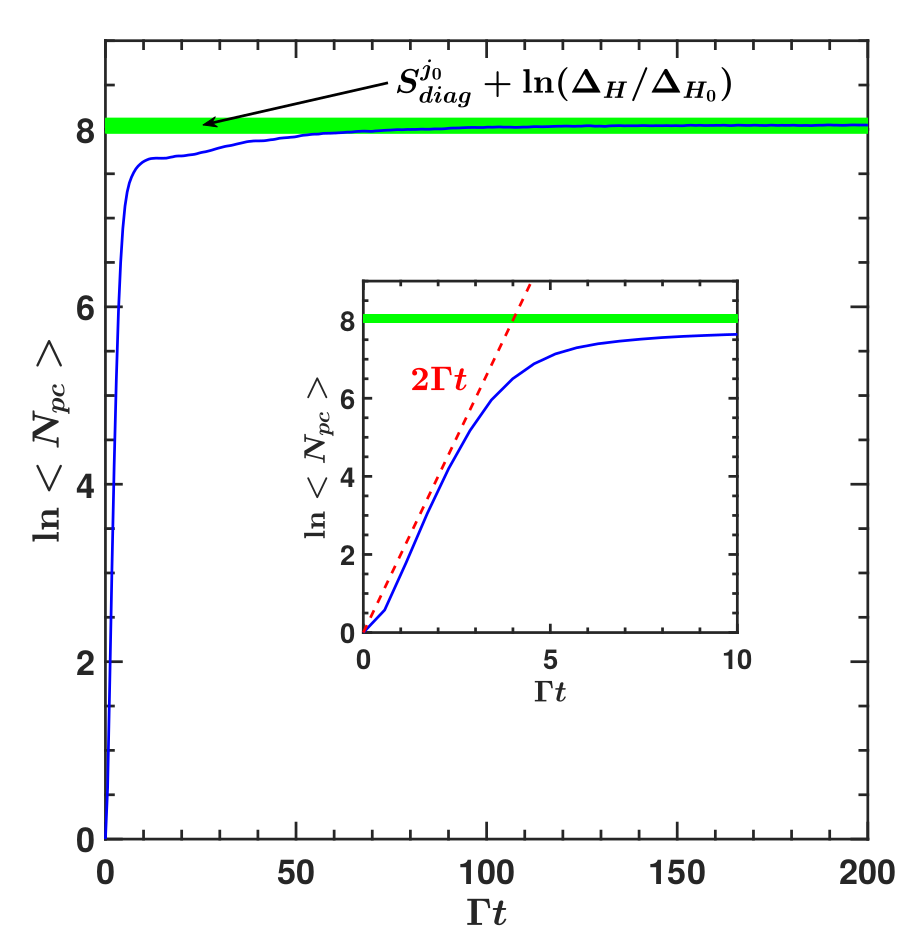}
	\caption{Time dependence of $\ln \langle N_{pc} \rangle$ for the TLL model. On x-axis we put the dimensionless time $\Gamma t$ where $\Gamma$ is the  width of the LDoS averaged over the degenerate initial states. The average has been done over all initial degenerate states having the same unperturbed energy $E_0$ close to the band center (with $j_0 \approx 4050$). Inset: the early stage of the evolution of $\ln \langle N_{pc} \rangle$, in comparison with linear dependence $2\Gamma t$ (red dashed line). Here we considered $N=9$ particles in $\ell=13$ momentum levels, $n/g=0.5 $, for a fixed total momentum $\mathcal {M}=1$ (the matrix size is $8122$). Horizontal green thick line represents the normalized diagonal entropy $S_{diag}+\ln (\Delta_H/\Delta_{H_0})$, the thickness corresponds to one standard deviation due to different initial states. In computing $\Delta_H$ and $\Delta_{H_0}$ we excluded a number of energies close to the band edges.
}
\label{fig:pr}
\end{figure}

Our numerical study of the quench dynamics of the TLL model demonstrates that $ N_{pc}(t) $ oscillates in time in the MF regime  ($n/g \gg 1 $) as shown in Fig.\ref{fig:wpr}. In contrast, it grows exponentially fast in the TG regime ($n/g \lesssim 1 $, see Fig.~\ref{fig:pr}),  after a short time where, due to the standard perturbation theory, the time-dependence is quadratic in time. This exponential increase lasts up to some critical time $t_s$ after which a clear saturation of $N_{pc}$ emerges, together with irregular fluctuations around its mean value. In order to reduce these fluctuations which are due to different initial conditions (various values of $j_0$), we have performed the average $\langle N_{pc} \rangle$ over all those initial states with the same unperturbed (degenerate) energy. 

\begin{figure}[t]
	\centering
	\includegraphics[width=12cm]{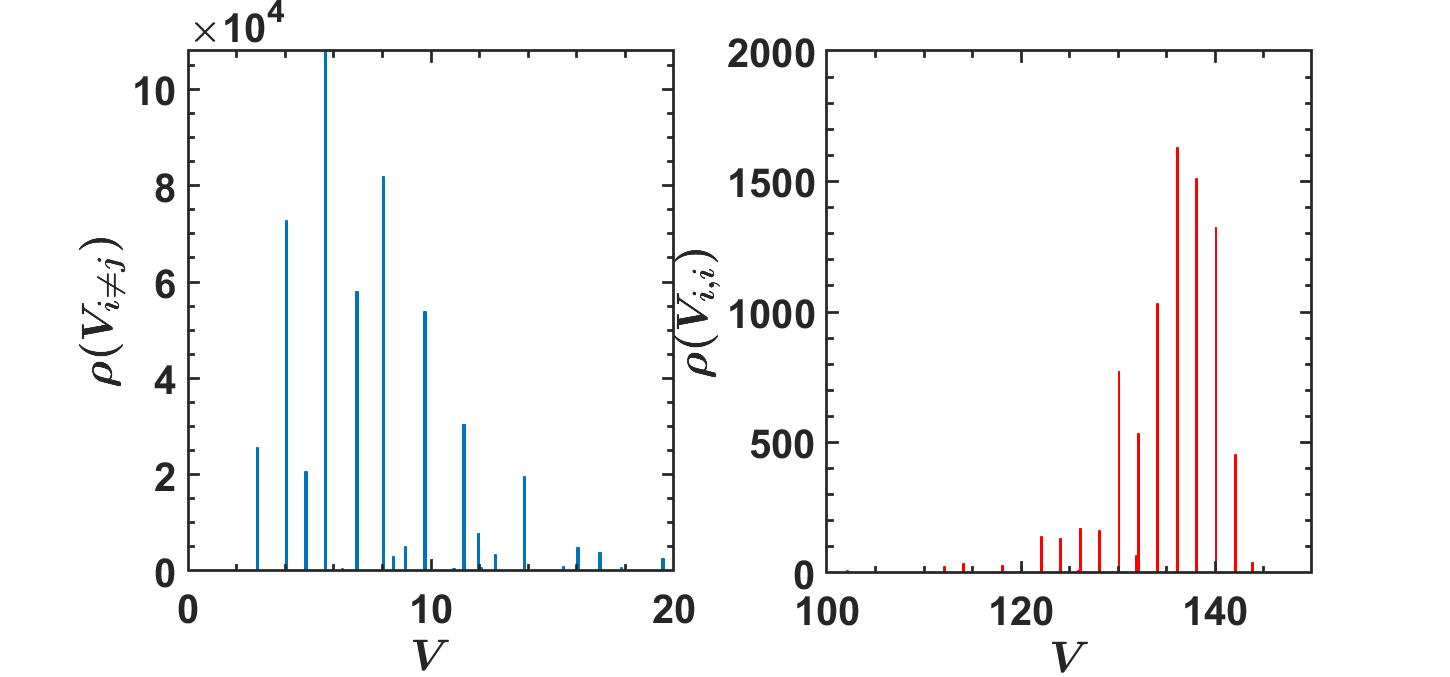}
	\caption{Truncated LL model (TLL). Left panel : distribution of off-diagonal elements $\rho(V_{i,j})$ for the Hamiltonian matrix  Eq.~\ref{eq:ham}. Right panel : distribution of diagonal matrix elements $\rho(V_{i,i})$. Here we considered 
$N=9$ particles in $\ell=13$ momentum levels and $g=1$. 
}
\label{fig:vdis}
\end{figure}

First, the numerical data clearly manifest a very good correspondence to the analytical estimates of the exponential increase of $ N_{pc}(t) $ occurring for $t \ll t_s$. Second, the relaxation time $t_s$ roughly corresponds to the estimate $t_s \approx N/\Gamma$ (specifically, to $\Gamma t_s \approx N$ with $N=9$). Finally, the relation (\ref{relation}) between the diagonal and thermodynamic entropies holds with a very good accuracy. All these results are highly non-trivial since they are obtained for the TLL model which is non-random. Specifically, one can show that the non-diagonal matrix elements of the total Hamiltonian are strongly correlated. For the case shown in Fig.\ref{fig:pr} diagonal and off-diagonal Hamiltonian matrix elements take the values from a set with very few elements, as one can see from the plots shown in Fig.\ref{fig:vdis}.

\section{Summary} 

We have studied two different models, the TBRI model with random two-body interaction and the truncated LL model with a finite number of particles in  a finite number of momentum states which is originated from the completely integrable Lieb-Liniger model. By studying the quench dynamics in the region where the many-body eigenstates can be considered as strongly chaotic, we have found that the time evolution of both models is quite similar to that recently found for the random TBRI, as well as for the deterministic XXZ model of interacting 1D spins-1/2 \cite{SBI12,SBI12a}. Specifically, the number of components in the wave packets evolving in the Hilbert space, increases exponentially in time with a rate given by twice the  width of the LDoS related to the initial state of the unperturbed many-body Hamiltonian $H_0$. This growth lasts approximately until the saturation time $t_s$ which is much larger than the inverse width of the LDoS characterizing the initial decay of the survival probability. Both the rate of the exponential increase and the characteristic time $t_s$ for the onset of equilibration, nicely correspond to our semi-analytically estimates. 

By studying the process of relaxation of the system to the equilibrium, we have discovered a remarkable relation between the thermodynamic entropy $S_{th}$ (defined in terms of the number of principal components $N_{pc}$ in the wave function) of the system {\it after the equilibration}, and the diagonal entropy $S_{diag}$ related to an {\it initial many-body state}. This relation (\ref{relation}) establishes a direct link between statistical and thermodynamical properties, and seems to be generic, valid for both deterministic and random  many-body systems. Recently, another relation between the diagonal entropy and the GGE entropy was found in a fully integrable 1D Ising model in a transverse magnetic field \cite{KBC14,PVCR17}. Thus, a further study of the relevance of the diagonal entropy to the thermodynamic properties of many-body systems seems to be very important, especially, in view of possible experimental studies of the onset of thermalization in isolated systems. 
 
{\em Acknowledgements.}--
F.B. acknowledges support by the Iniziativa Specifica I.N.F.N.-DynSysMath and FMI acknowledges financial support from CONACyT (Grant No. 286633).


\begin{thebibliography} {99}
\bibitem{new} The details will be published elsewhere.
\end{thebibliography}
\end{document}